\documentclass[a4paper,11pt]{article}
\pdfoutput=1 

\usepackage{jcappub}

\usepackage[T1]{fontenc} 

\usepackage{amsmath,amssymb,amsfonts}
\usepackage{bm, slashed, soul}
\usepackage{color,graphicx}
\usepackage{hyperref}

\usepackage{xcolor}
\definecolor{red}{rgb}{0.9, 0,0}

\def\tilB{{\tilde{B}}}
\def\tilL{{\tilde{\Lambda}_{QCD}}}

\def\tilG{{\tilde{\Gamma}}}
\def\tT{{\tilde{T}}}

\def\tp{\tilde{p}}
\def\tn{\tilde{n}}
\def\tpi{\tilde{\pi}}
\def\tD{\tilde{D}}

\title{\boldmath Asymmetric Twin Dark Matter}

\author{Marco Farina}

\affiliation{Department of Physics, LEPP, Cornell University, Ithaca, NY 14853, USA}
\emailAdd{mf627@cornell.edu}

\abstract{We study a natural implementation of Asymmetric Dark Matter in Twin Higgs models. The mirroring of the Standard Model strong sector suggests that a twin baryon with mass around 5 GeV is a natural Dark Matter candidate once a twin baryon number asymmetry comparable to the SM asymmetry is generated. We explore twin baryon Dark Matter in two different scenarios, one with minimal content in the twin sector and one with a complete copy of the SM, including a light twin photon. The essential requirements for successful thermal history are presented, and in doing so we address some of the cosmological issues common to many Twin Higgs models. The required interactions we introduce predict signatures at direct detection experiments and at the LHC.
}

\begin{document}
\maketitle
\flushbottom

\section{Introduction}
There is overwhelming evidence for the presence of Dark Matter in the Universe, however its nature is yet to be unveiled. One particularly suggestive observation is that $\Omega_{\text{DM}} \simeq 5 \Omega_B$, raising the question of whether a common origin for the baryon and DM abundances is possible. One answer to this question is the Asymmetric Dark Matter (ADM) picture~\cite{Nussinov:1985xr,Chivukula:1989qb,Barr:1990ca,Kaplan:1991ah,Hooper:2004dc,Kaplan:2009ag} (for recent reviews see~\cite{Davoudiasl:2012uw,Petraki:2013wwa,Zurek:2013wia}), which assumes the DM density to be determined by an asymmetry $n_{\text{DM}}$ in the same way the baryon number asymmetry $n_B$ sets the visible matter density. In general the abundances are related by
\begin{equation}
  \frac{\Omega_{\text{DM}}}{\Omega_B} = \frac{n_{\text{DM}}}{n_B} \frac{m_{\text{DM}}}{m_N} \,\, ,
\end{equation}
where $m_{\text{DM}}$ is the DM mass and $m_N$ is the nucleon mass. If the two asymmetries are generated by the same mechanism or if one is responsible of the other then we expect $n_{\text{DM}} \sim n_B$. It follows that $m_{\text{DM}}\sim 5 m_N$ and the question to address is why this mass ratio is close to unity.

The discovery of the Higgs boson and the lack of new physics signals at the LHC made Naturalness and the hierarchy problem a critical issue more than ever before. In search of solutions, renewed attention has been recently brought to Twin Higgs (TH) models~\cite{Chacko:2005pe,Chacko:2005un,Chacko:2005vw,Burdman:2014zta,Barbieri:2005ri} that solve the little hierarchy problem by introducing a copy of the SM. The mirroring of the SM Lagrangian is due to a $Z_2$ symmetry. The scalar potential has an accidental global $SU(4)$ symmetry which protects the Higgs potential from quadratic corrections. The Higgs is indeed a pseudo-Nambu-Goldstone boson and it is naturally light even after symmetry breaking, which happens at loop level. For Naturalness reasons the order parameter $f$ of the $SU(4)$ breaking is expected to be $f \sim \mbox{few } v$, with $v=246$ GeV the SM Higgs vev.

Twin Higgs is a natural environment in which to implement the ADM idea. It is straightforward to draw a comparison between baryons and twin baryons and to consider the latter as DM candidates. Both are stable thanks to conservation of the SM (twin) baryon number $B$ ($\tilB$). Approximate $Z_2$ symmetry suggests $n_{\text{DM}} \sim n_B$, for example by having only the combination $B-\tilB$ conserved above the scale $f$, and $m_{\text{DM}}\sim m_N$, with the second easily perturbed by the breaking of $Z_2$ below $f$. Motivated by this tantalizing observation we study the viability of twin baryons as ADM.

Related work~\cite{Garcia:2015loa,Craig:2015xla,Garcia:2015toa} on both thermal and asymmetric DM has been recently carried out in the context of Fraternal TH models~\cite{Craig:2015pha}, in which only the third generation is mirrored. Here instead we focus on the conventional TH scenario with a copy of all generations.

\section{Twin sector properties}\label{Setup}

Our starting point is a Twin Higgs model with an effective cutoff $\Lambda \sim 4 \pi f \sim 5-10$  TeV, where again $f \simeq \mbox{few } v$ for Naturalness. We will suppose that there is a copy of each SM quark charged under twin $\tilde{SU(3)}$ QCD and twin $\tilde{SU(2)}$ weak gauge symmetries, whose couplings are expected to be close to the SM counterparts $\tilde{g}_i (\Lambda) \simeq g_i(\Lambda) $ (from now on all tilded symbols refer to twin sector quantities and fields). Hypercharge is not necessarily gauged, as recently shown in~\cite{Craig:2015pha,Barbieri:2015lqa}, and eventually the twin photon is either massless or very heavy $\tilde{m}_\gamma \simeq \Lambda$. The presence of leptons in the spectrum is not strictly necessary and we will comment on different possibilities in the next section.

Naturalness constrains the twin top and bottom Yukawas, with $\tilde{y}_t$ and $\tilde{y}_b$ expected to fall respectively within $1\%$ and $10\%$ of the SM values, so from now on we fix $\tilde{y}_t=y_t$.
While in general $\tilde{y}_i \simeq y_i $ is still expected from $Z_2$ symmetry, naturalness does not impose any bound on the first two generations. Hence large deviations can be possible, but we will not consider this in the following. Similar considerations apply to the leptons when present; either they will be assumed to have mass $\tilde{m}_\ell \simeq f/v~m_\ell$, or to be integrated out $\tilde{m}_\ell \gg v$.

As for the interactions between the two sectors any TH model contains Higgs interactions
\begin{equation}
  \mathcal{L} \supset y_f h \bar{f} f - \frac{y_f}{\sqrt{2}f} h^\dagger h  \bar{\tilde{f}} \tilde{f} \,\, ,
\end{equation}
where $h$ is the SM-like Higgs and its interactions with twin particles are those responsible for  cancellation of quadratic divergences. At low energy they induce an effective operator
\begin{equation}
  \mathcal{O}_h \equiv \frac{m_i \tilde{m}_{j}}{m_h^2 f^2} ( \bar{f}_i f_i) ( \bar{\tilde{f}}_j \tilde{f}_j) \,\, .
  \label{eq:HiggsPortal}
\end{equation}
Higher dimensional operators will also be introduced in the following, with an expected effective scale $M \sim 4 \pi f$. In particular we have in mind recent realization of TH in the composite holographic Higgs framework~\cite{Geller:2014kta,Barbieri:2015lqa,Low:2015nqa}, in which effective operators are generated by the strong sector and by integrating out heavy resonances.

Finally the twin sector respects the same accidental global symmetries as the SM: twin baryon number, lepton number and charge. We will neglect possible small breakings generated by higher dimensional operators and consider stable each of the lightest particles carrying global charges.

\subsection{Twin Dark Matter}

Due to conservation of twin baryon number $\tilB$, as already explained, twin proton $\tp$ and eventually $\tn$ are stable on cosmological time scales. If in the early Universe comparable asymmetries are generated then $n_B \simeq n_{\tilB}$ and
\begin{equation}
  \frac{\Omega_\tilB}{\Omega_B} \simeq \frac{\tilde{m}_N}{m_N} \,\, ,
\end{equation}
where again $m_N$ ($\tilde{m}_N$) is the (twin) nucleon mass and thus
\begin{equation}
  \tilde{m}_N \simeq 5~m_N \,\,.
  \label{eq:Rmass}
\end{equation}
As in the SM $\tilde{m}_N$ is determined by the QCD confinement scale $\tilL$, unless deviations of various order of magnitude are allowed in the Yukawas. Then satisfying Eq.~(\ref{eq:Rmass}) requires
\begin{equation}
  \tilL \simeq 5~\Lambda_{QCD} \,\, .
\end{equation}
How likely is it to happen? The running of $\alpha_s$ plays a crucial role. First the different masses of the twin quarks change the thresholds in the $\beta$ function. It is also possible to directly introduce a small difference $\delta \alpha_s=\tilde{\alpha}_s(\Lambda)-\alpha_s(\Lambda)$, for instance this can be due to threshold effects in composite TH realizations, in which heavy resonances are expected to have different representations under the twin and the SM gauge groups.

\begin{figure}
\begin{center}
\includegraphics[width=0.5\textwidth]{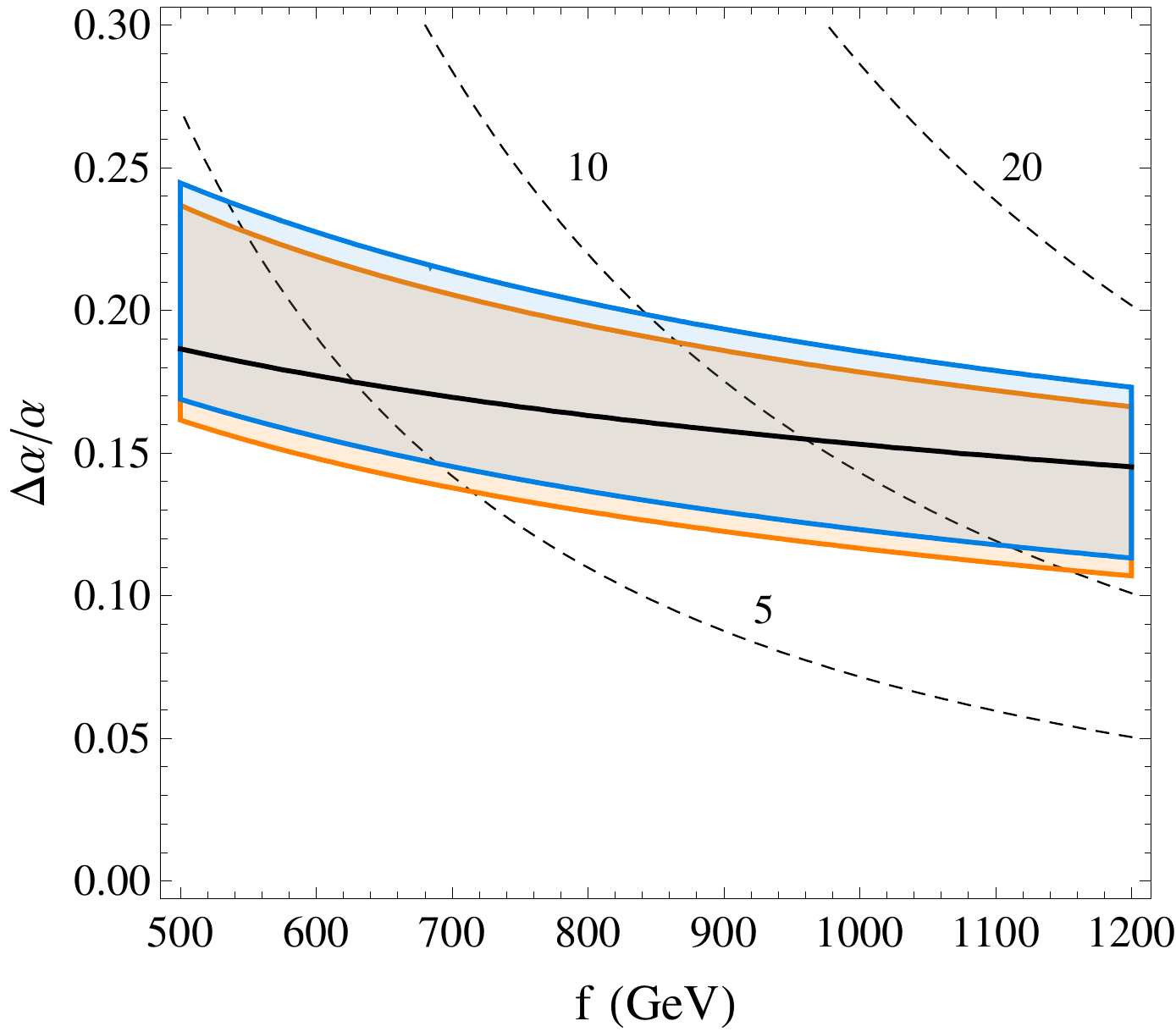}
\caption{The region of parameter space where $6> \tilL/\Lambda_{QCD} >4$ as a function of $f$ and $\delta \alpha_s=\tilde{\alpha}_s(\Lambda)-\alpha_s(\Lambda)$. In blue the case in which $\tilde{y}_q=y_q$, in orange the case $\tilde{y}_q=2 y_q$. The solid line corresponds to $\tilL/\Lambda_{QCD}=5$ computed using the approximate function defined in Eq.~(\ref{eq:rLQCD}). Dashed lines depict the fine tuning associated with two loops strong contributions to the Higgs mass, for FT values $\Delta=5,~10,~20$.
\label{fig:LQCD}}
\end{center}
\end{figure}

We compute the confinement scale of twin QCD by running $\tilde{\alpha}_s$ down from $4 \pi f$, with the $\beta$ function computed for 6 flavours of mass $\tilde{m}_q = f/\sqrt{2}~\tilde{y}_q$. This leaves $\delta \alpha_s$, $f$ and  rescalings of all Yukawas (apart the $\tilde{y}_t$) as variables. The results are presented in Fig.~\ref{fig:LQCD} which shows how $\tilL/\Lambda_{QCD}\simeq 5$ is a typical result once a $O(10\%)$ splitting of the UV coupling constants is introduced. The effect can be explained by a simple approximate equation obtained by fixing $\tilde{y}_i=y_i$
\begin{equation}
  \frac{\tilL}{\Lambda_{QCD}} \simeq \left(\frac{f}{v}\right)^{2/9} \text{Exp}\left[\frac{2 \pi}{9} \left(\frac{1}{\alpha_s(\Lambda)}-\frac{1}{\tilde{\alpha}_s(\Lambda)}\right)\right] \,\, .
\label{eq:rLQCD}
\end{equation}
The first term accounts for the difference in the quark mass thresholds and its evident suppression explains the negligible sensitivity to the choice of $f$ and $\tilde{y}$. On the other hand the initial condition is exponentially amplified by the running. We also show the fine tuning (FT) contours given by two loop contribution to the Higgs mass~\cite{Craig:2015pha}
\begin{equation}
  \delta m_h^2 \simeq \frac{3 y_t^2 \Lambda^2}{\pi^3}(\alpha_s-\tilde{\alpha}_s) \,\,,
\end{equation}
which is of the same order of the standard $FT\sim (f/v)^2$. Then $f/v \lesssim 5$ is expected to avoid FT larger than few $\%$.

Having successfully accounted for the DM to baryon mass ratio, how can their number density be related?
Following the ADM picture~\cite{Kaplan:2009ag} some mechanism connecting the baryon asymmetries in the two sectors should be efficient in the early Universe. First consider an effective operator connecting SM and twin singlets, $Z_2$ invariant and preserving only $B-\tilde{B}$
\begin{equation}
  \mathcal{O}_{B\tilde{B}} \equiv \frac{g_{ijk}^{lmn}}{M^5} (u_i d_j d_k) (\tilde{u}_l \tilde{d}_m \tilde{d}_n) \,\, ,
\end{equation}
where again $M \sim 4 \pi f$ and all fields are right handed. The operator flavour structure, embedded in the coupling $g_{ijk}^{lmn}$, has been introduced to evade constraints on DM lifetime. A structure of the form $g_{ijk}^{lmn} \propto y_i y_j ...$, as in partial compositeness,  would be enough to avoid any bound and to ensure that the operator involving third generation quarks is in equilibrium at temperatures $f \lesssim T\lesssim4 \pi f$. Indeed for $g\sim1$ the operator is efficient above temperatures $T \sim M$ and, because it conserves $B-\tilde{B}$, it will convert any excess in $B$ into $\tilde{B}$ and viceversa so that it enforces $\tilde{n}_{B} \sim n_B$.

A different possibility is the sharing of asymmetries through non-perturbative effects, for example leptogenesis followed by generation of both $B$ and $\tilB$. This solution depends crucially on the UV completion of the model and it is beyond the scope of the present work.

We conclude, also anticipating the results in Section~\ref{sec:DM}, that our DM candidate is a twin neutron $\tilde{n}$ with mass $\tilde{m}_n \simeq 5$ GeV. The $\tilde{n}$ is typically stable and it is the main DM component due to charge neutrality of the Universe if the $\tilde{B}$ asymmetry is the only asymmetry generated in the twin sector.  We will study its experimental signatures in Section~\ref{exp}. Let us remark that DM self interaction scales like $(\Lambda_{QCD}/\tilL)^3$ resulting in a cross section $\sigma_{\tilde{n}}/\tilde{m}_n \simeq 0.25~\text{cm}^2~\text{g}^{-1}$ which is below the current bound $\sigma_{\tilde{n}}/\tilde{m}_n\lesssim 0.5~\text{cm}^2~\text{g}^{-1}$~\cite{Zavala:2012us,Harvey:2015hha}.

\section{Thermal History }\label{sec:DM}

A successful thermal history must address two main points: the absence of other, possibly overproduced, relics and the influence on BBN and CMB of additional relativistic degrees of freedom.

Notice that we cannot rely on low reheating temperatures. The Higgs portal of Eq.~(\ref{eq:HiggsPortal}) will always keep the two sectors in thermal equilibrium above temperature $T \simeq$ few GeV, while the DM (and baryon asymmetry) are expected to be generated at $T\gtrsim v$. As such we discard reheating as a solution and from now on we always consider the two sectors to be in equilibrium up to at least $T \simeq$ few GeV.

In the following we will present the thermal history, including a possible twin nucleosynthesis phase, for two scenarios differing in content of the twin sector:
\begin{itemize}
  \item Scenario A: a minimal scenario with only twin quarks.
  \item Scenario B: a complete copy of the SM content, including a massless twin photon.
\end{itemize}


\subsection{Scenario A}
We start with a minimal scenario with only the necessary degrees of freedom, which are the twin quarks. We consider either the twin photon heavy or the twin hypercharge not gauged. In the latter case there is no gauge anomaly and no leptonic field is required \footnote{It must be pointed out that at least one extra twin-$SU(2)$ doublet is necessary because of the global Witten anomaly\cite{Witten:1982fp}, which requires an even number of doublets. For simplicity such additional field is supposed to have mass above the $v$, for example $m=y f/\sqrt(2)$ with $y\sim 1$, and be negligible in our discussion.}, in the former all charged leptons and neutrinos are taken to be heavy enough to be integrated out. We will comment on this choice at the end of the section.

This scenario does not include any (nearly) massless particle and has no impact on the CMB. Only twin pions as possible symmetric relics are left to study. Their mass is given by
\begin{equation}
 \tilde{m}_\pi \simeq \sqrt{\frac{f}{v} \frac{\tilL}{\Lambda_{QCD}}}  m_\pi  \simeq 500~\text{MeV} \left(\frac{f}{700~\text{GeV}}\right)^{1/2} \,\, .
\end{equation}
and similarly their decay constant scales as $\tilde{f}_\pi \simeq \tilL/\Lambda_{QCD} f_\pi \simeq 450$ MeV.
The $\tpi^{\pm}$ are stable as they are the lightest charged particles. On the other hand $\tpi^{0}$ is a singlet and can decay to SM states, with its width $\tilG $ being the crucial parameter. We assume it to be induced by higher dimensional operators.

To study the cosmological history of the twin-pions we write the corresponding Boltzmann equations. We assume that operators responsible for $\tpi^0$ decay keep the twin pions in thermal equilibrium via the processes $\tpi~\mbox{SM} \rightarrow \tpi~\mbox{SM}$. Neglecting the twin baryons, the Boltzmann equations are
\begin{eqnarray}
    \dot{n}_0+ 3 H n_0 &=&  - \langle \Gamma \rangle   (n_0-\bar{n}_0)-2 \langle \sigma v  \rangle (n_0^2-n_\pm n_\pm) \,\, , \nonumber \\
\dot{n}_\pm+ 3 H n_\pm &=& -\langle \sigma v  \rangle (n_\pm n_\pm -n_0^2) \,\, ,
\label{eq:Boltzmann}
\end{eqnarray}
where $n_0$ and $n_\pm$ are the number densities of $\tilde{\pi}^0$ and $\tilde{\pi}^\pm$ respectively, and $\bar{n}$ represents the equilibrium distribution. Here, angle brackets represents thermal averages and $\sigma$ is the $\pi \pi$ scattering cross section
\begin{equation}
  \sigma(\pi\pi \rightarrow \pi \pi) = \frac{1}{16 \pi \tilde{f}^4_\pi} \frac{(s-\tilde{m}^2_\pi)^2}{s} \sqrt{1-\frac{4\tilde{m}^2_\pi}{s} } \,\, .
  \label{eq:sigma}
\end{equation}
Upon inspection of Eq.~(\ref{eq:Boltzmann}) it is clear that if $\tilG \gg 2 H(T=\tilde{m}_\pi)$ the decay is efficient in keeping $\tpi$ in equilibrium and no relic density will be produced. For $\tilG \ll 2 H(T=\tilde{m}_\pi)$,  equilibrium is not maintained and the $\tpi$ will become overabundant, with each degree of freedom having a freeze-out density
\begin{equation}
  \tilde{Y}=0.278 / g_{*S}(T=\tilde{m}_\pi) \,\,.
\end{equation}
After freeze-out the term proportional to $\bar{n}$ in Eq.~(\ref{eq:Boltzmann}) drops to zero. The first term on the right hand side describes the decay of $\tpi_0$, depleting their number, while the second term replenishes it as long as $\sigma \tilde{n} \gg \tilG$. Above a certain temperature $\tilde{T}$ the scattering will be efficient, $n_\pm$ will track $n_0$ and the total number density will decrease as $\tilde{n}_i(T) = \tilde{n}_i(T_{fo})\text{Exp}[-\Gamma/(6 H(T))]$. On the other hand below $\tilde{T}$ the $\tpi^{\pm}$ number density will freeze-out while all $\tpi^0$ will eventually decay. A rough estimate of $\tilde{T}$ can be computed by imposing
\begin{equation}
\frac{\langle \sigma v \rangle s\tilde{Y} \text{Exp}[-\Gamma/(6 H)]}{H} \Big|_{T=\tilde{T}} \simeq 1 \,\, ,
\label{eq:tildeT}
\end{equation}
where $s(T)$ is the entropy density. The final density of $\tpi_\pm$ is then given by
\begin{equation}
  \Omega_{\pi} h^2 \simeq 2 \left(\tilde{m}_\pi \tilde{Y} (s_0/\rho_c) \right) \text{Exp}\left[-\Gamma/(6 H(\tilde{T}))\right]  \,\, .
\label{eq:OmegaApprox}
\end{equation}
An upper bound on the $\tpi$ width can be estimated by plugging the solution of Eq.~(\ref{eq:tildeT}) in Eq.~(\ref{eq:OmegaApprox}). Imposing a conservative limit $\Omega_{\pi} h^2 <1/10~\Omega_{\text{DM}}^{\text{obs}} h^2 \simeq 0.012$, we obtain
\begin{equation}
  \tilG \gtrsim 10^{-25} \mbox{ GeV} \,\,.
  \label{eq:boundG1}
\end{equation}

Moreover, if the $\tpi^0$ lifetime is too long its decays could inject entropy during BBN. Thus we require $1/\tilG \lesssim 1$ s, corresponding to
\begin{equation}
 \tilG \gtrsim 6.6 \times10^{-25} \mbox{ GeV} \,\,.
\label{eq:boundG2}
\end{equation}
The bound of Eq.~(\ref{eq:boundG2}) automatically satisfies Eq.~(\ref{eq:boundG1}) and suggests that no significant relic density of $\tpi^\pm$ is left and DM can be considered as solely composed of twin baryons. It is interesting to notice that such a late decay could be linked, and possibly solve, the cosmological lithium problem \cite{Pospelov:2010cw}.

Finally it is possible that for sufficiently long lived pions the Universe enters an early period of matter domination with subsequent reheating once the pions decay. This would dilute both baryon and DM asymmetries. While irrelevant for present discussion, it is an important point to address in any UV completion.

If twin hypercharge is gauged the $\tpi^0$ decays in SM via anomaly and $\gamma\text{-}\tilde{\gamma}$ mixing \cite{Essig:2009nc} and the width is\footnote{We are thankful to Tongyan Lin for pointing out an error in Eq.~(\ref{eq:gammagamma}) of the original version of the paper} 
\begin{equation}
  \tilde{\Gamma} = \frac{\tilde{\alpha}^2}{64 \pi^3}\frac{\tilde{m}_\pi^3}{\tilde{f}_{\pi}} \, \left(\frac{1}{4\pi} \alpha \epsilon^2 \frac{\tilde{m}_\pi^4}{\tilde{m}_{\gamma}^4} \right)^2 \,\,.
  \label{eq:gammagamma}
\end{equation}
where $\tilde{m}_{\gamma} \approx$ TeV is the twin photon mass.
However the condition $\tilG \gg 2 H$ is never satisfied when compared with the experimental constraint of $\epsilon \lesssim 10^{-1}\text{-}10^{-2}$ ~\cite{Davidson:2000hf,Vogel:2013raa}. Alternatively, and in the case of hypercharge not gauged the $\tpi^0$ decay can be obtained by introducing higher dimensional operators. In particular, an axial current term $(1/M^2) ( \bar{q} \gamma_\mu \gamma_5 q) ( \bar{\tilde{q}} \gamma_\mu \gamma_5 \tilde{q})$ induces a decay through mixing with the SM mesons. Notice that such operators involve isospin singlets and so we include isospin breaking effects in the form of the $\tpi\text{-}\tilde{\eta}'$ mixing angle $\sin{\theta} \sim 10^{-2}$. The width roughly scales as
\begin{equation}
  \tilde{\Gamma} \simeq  \frac{f^2_\pi \tilde{f}^2_\pi}{M^4} \frac{\tilde{m}^3_\pi}{m^3_X} \sin^2 {\theta} \, \Gamma_X \,\,,
\end{equation}
where the first factor accounts for the mixing between the pion and an $X$ meson, while the mass ratio accounts for the phase space difference. To avoid additional isospin breaking suppressions from the SM sector we consider mixing with the $\eta'$ meson and so Eq.~(\ref{eq:boundG2}) translates into
\begin{equation}
  M \lesssim 2.5 \mbox{ TeV }\left(\frac{\tilde{m}_\pi}{1 \mbox{ GeV}}\right)^{3/4} \,\,.
  \label{eq:MBound}
\end{equation}
It is clear that for $\tilde{m}_\pi \lesssim m_{\eta'} =958$ MeV the required scale would be too small for any reasonable model. The price to pay for heavy $\tpi$ is in fine tuning as $\tilde{m}_{\pi}\propto \sqrt{f}$ and thus we expect $f/v \gtrsim 10$ (or drastic changes in the Yukawas).

Another possibility is a 4-fermion interaction between twin quarks and SM leptons. The width is
\begin{equation}
\tilde{\Gamma}(\tilde{\pi}^0 \rightarrow \ell \bar{\ell}) = \frac{1}{32 \pi M^4} m_\ell^2 \tilde{m}_\pi \tilde{f}^2_\pi  \sin^2{\theta} \,\left(1-\frac{4 m_\ell^2}{\tilde{m}^2_\pi}\right)^{1/2}  \,\,,
\end{equation}
where $m_\ell$ appears because of helicity suppression and decay to muons is the dominant channel. The bound from Eq.~(\ref{eq:boundG2}) gives
\begin{equation}
  M \lesssim 6 \mbox{ TeV }\left(\frac{\tilde{m}_\pi}{500 \mbox{ MeV}}\right)^{3/4} \,\,,
\end{equation}
which is less stringent than Eq.~(\ref{eq:MBound}), but the presence of such operators is less easily justifiable.

What if twin leptons are added? Apparently the $\tpi^\pm$ stability problem would be eliminated if $\tpi^\pm \rightarrow \tilde{l}^\pm \tilde{\nu}$ decay is allowed. However $\tilde{\ell}$ annihilate only through weak interactions to $\tilde{\nu}$, and a modified version of the Lee-Weinberg bound~\cite{Lee:1977ua} then requires $\tilde{m}_{\ell} >$ few GeV to avoid overproduction. It would clearly reintroduce the stable $\tpi^{\pm}$ problem. On the other hand, it is an interesting observation that a heavy $\tilde{\ell}$ could be a WIMP candidate, a possibility recently explored in~\cite{Garcia:2015loa,Craig:2015xla}.
The presence of light neutrinos does not significantly change the $\tilde{\pi}^0$ decay width, cause of the mass suppression from chirality flip, and bounds from CMB could be possibly constraining. We do not pursue this direction any further.

Finally let us briefly comment on the possibility of twin nucleosynthesis, which could involve $\tp$ and $\tn$ once the temperature of the Universe is around the twin deuteron $\tilde{D}$ binding energy. Notice that $\tn$ is stable thanks to phase space. The fusion processes can proceed only with a pion in the final state with $\tp \tn, \tp \tp,\tn \tn \rightarrow \tD \tpi^{0,\pm}$. At temperatures much above the binding energy the ratio of nucleons to nuclei is set by the high entropy of the Universe and the abundance of nuclei is negligible. In particular, this is valid at $T \lesssim \tilde{m}_\pi$ when the fusion is not efficient thanks to Boltzmann suppression. We conclude that DM is entirely composed of $\tilde{n}$ and possibly $\tilde{p}$ (if hypercharge is not gauged).


\subsection{Scenario B}

Now we consider a complete copy of the SM and again assume $\tilde{y}_i\simeq y_i$.
Various experimental constraints~\cite{Davidson:2000hf,Vogel:2013raa} require a small photon-twin photon kinetic mixing $\epsilon \lesssim 10^{-9}$, which could be possible if mixing arises at 4-loop or higher order. A second constraint comes from the number of effective neutrino species, the recent Planck measurements~\cite{Planck} give $\Delta N_{eff}= 0.11 \pm 0.23$. The number of relativistic degrees of freedom is defined
\begin{equation}
g_*(T)= \sum_{i=\text{bosons}} g_i \left(\frac{T_i}{T}\right)^4+ \frac{7}{8}\sum_{i=\text{fermions}} g_i \left(\frac{T_i}{T}\right)^4 \,\,,
\end{equation}
where $g_i$ accounts for the internal degrees of freedom. In general $g_*$  in the twin sector is comparable to the one in the SM. Above the decoupling temperature the only difference is given by the larger mass of twin particles, which become non relativistic at a higher temperature. After decoupling the different temperature of the two sectors must be taken into account.

If the twin neutrinos are light, the contribution of $\tilde{\nu }$ and $\tilde{\gamma}$ is equal to the SM one, $g_*=3.36$, rescaled to take into account the different temperatures of the two sectors
\begin{equation}
  \Delta N_{eff} \simeq  \frac{3.36}{0.45} \left(\frac{\tT}{T}\right)^4 \,\,,
  \label{eq:Neff1}
\end{equation}
where the normalization factor corresponds to $g_*=0.45$ of one SM neutrino species. By conservation of entropy the temperatures ratio is
\begin{equation}
   \frac{\tT}{T}  \simeq \left(\frac{\tilde{g}_*(T_d)}{g_*(T_d)} \right)^{1/3} \,\,,
\end{equation}
with $T_d$ the temperature of decoupling of the two sectors. If the decoupling happens at the freeze-out of Higgs portal interactions, then $\tilde{g}_*(T_d) \sim g_*(T_d)$ and $\Delta N_{eff} \sim 7$, which is ruled out. The best possibility is to have the sectors decouple between the two QCD phase transitions when $g_*(T_d)\sim 61.75$ while $\tilde{g}_*(T_d) \sim 10.75$. Thus we obtain $\Delta N_{eff} \sim 0.7$, which is however still in tension with Planck measurements.

The most natural solution is to decouple the $\tilde{\nu}$ by lifting their masses, for instance by a suitable choice of the twin seesaw scale. The effect is two-fold: increasing the entropy ratio at decoupling, $\tilde{g}_*(T_d) \sim 5.5$, and decreasing the total number of relativistic species as can be readily seen by a modified version of Eq.~(\ref{eq:Neff1})
\begin{equation}
  \Delta N_{eff} \simeq  \frac{2}{0.45} \left(\frac{\tT}{T}\right)^4 \simeq 0.17 \,\,,
  \label{eq:Neff2}
\end{equation}
which is compatible with present constraints. We conclude that the two sectors must decouple between the two phase transitions with $0.15 \mbox{ GeV} \lesssim T_d \lesssim 1$ GeV. How can this be achieved? Additional interactions between the two sectors must be introduced and $\tilde{e}$ should be involved. For a 4-fermion operator suppressed by a scale $M$ the decoupling temperature is roughly given by the condition
\begin{equation}
  \frac{T_d^5}{M^4}= H(T_d) \,\,.
  \label{eq:TdA}
\end{equation}
Then the required scale is $7 \mbox{ TeV} \lesssim M \lesssim 30$ TeV. We stress that the above results are independent of $f$. The estimate only involves $\tilde{m}_e$, which is few order of magnitudes lower than $\Lambda_{QCD}$ for $\tilde{y}_e \simeq y_e$, and thus our conclusions are not sensitive to $f$.

If the neutrinos are integrated out, then each leptonic species is stable due to individual lepton number conservation and could in principle be a thermal relic. However, the $\tilde{\ell} \tilde{\ell} \rightarrow \tilde{\gamma} \tilde{\gamma}$ annihilation channel is efficient enough to eliminate any symmetric component~\cite{Berger:2008ti}.

In this scenario $\tpi^{\pm}$ as well as $\tn$ are stable due to phase space, while $\tilde{\pi}_0$ will decay to $\tilde{\gamma}\tilde{\gamma}$ through the twin chiral anomaly. The relic density of $\tpi^{\pm}$ is negligible, as shown in the previous section.

What about nucleosynthesis? First consider the case in which the only asymmetry generated in the twin sector is in $\tilB$. Then charge neutrality of the Universe imposes that only $\tilde{n}$ are generated after the twin QCD phase transition. Twin neutrinos are heavy and so weak processes of the form $\tilde{e}^+ \tn \leftrightarrow \tilde{\nu} \tp$ are suppressed. We are left with fusion processes $\tn \tn \rightarrow \tD \tpi^-$ and as in the previous section we conclude that DM is completely composed of $\tn$.

On the other hand, suppose a $\tilde{L}$ asymmetry is generated in the early Universe as well. Now $\tilde{p}$ and $\tilde{e}$ are also present, and by charge neutrality $\tilde{n}_p =\tilde{n}_e$. So the fusion process $\tp \tn \rightarrow \tD \gamma$ is open and twin atoms will be formed. The ratio of different nuclei is set by the initial leptonic asymmetry and we leave a detailed study of this case for future work.


\section{Experimental signatures}\label{exp}

\begin{figure}
\begin{center}
\includegraphics[width=0.5\textwidth]{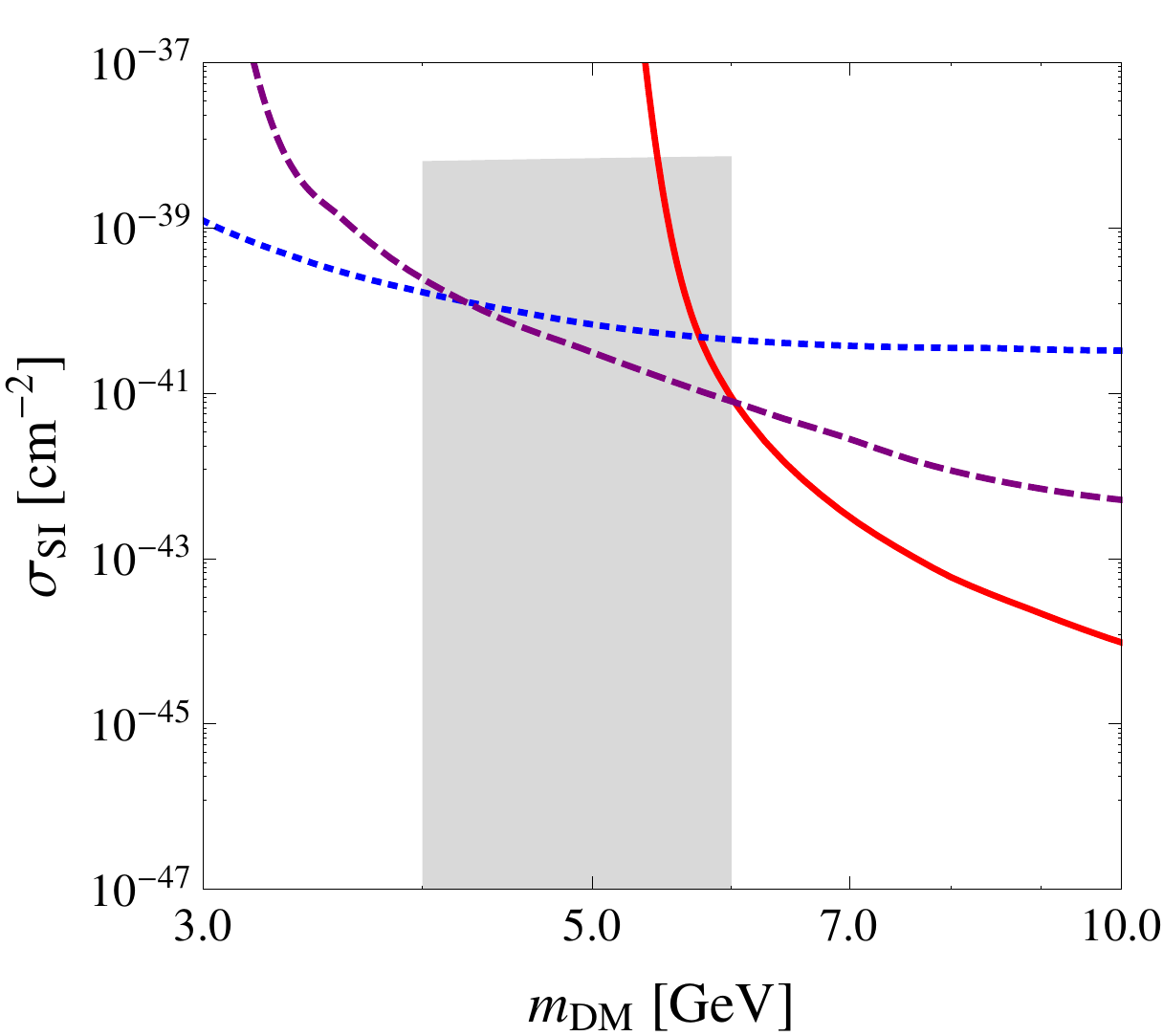}
\caption{DM direct detection parameter space. The shaded region corresponds to DM-nucleon spin independent cross sections mediated by the operator $\mathcal{O}_1$ for $6> m_{\text{DM}} >4$ GeV and $M > 1$ TeV. The lines correspond to the upper bounds from current DM direct detection experiments: CMDSlite~\cite{CMDSlite} (dotted blue), SuperCDMS~\cite{SuperCDMS} (dashed purple), LUX~\cite{LUX} (solid red).
\label{fig:DirectDetection}}
\end{center}
\end{figure}

The experimental signatures predicted by the model fall in two main categories: DM direct detection and possible collider signals. Both of them are crucially dependent on the higher dimensional operators that we have introduced in Section~\ref{sec:DM}. In general the presence of 4-fermion operators induce a DM-nucleon scattering cross section. For simplicity we consider an operator of the form
\begin{equation}
   \frac{b_q \tilde{b}_q}{M^2} ( \bar{q} \gamma_\mu q) ( \bar{\tilde{q}} \gamma_\mu \tilde{q}) \,\, .
\end{equation}
As previously explained we assume that DM is in the form of twin neutron with mass $\tilde{m}_n \simeq 5$ GeV. The spin independent cross section on a nucleon $N$ is
\begin{equation}
  \sigma_{SI}(\tilde{n}\text{-}N) = \frac{b_N^2 \tilde{b}_n^2}{\pi M^4}  \mu^2 \,\, ,
  \label{eq:sigmaSI}
\end{equation}
where $\mu$ is the reduced mass while $b_p=2b_u +b_d$, $b_n=b_u +2b_d$ and similarly for twin baryons. The parameter space with $M\lesssim3$ TeV is already probed by present experiments~\cite{CMDSlite,LUX,SuperCDMS} as shown is shown in Fig.~\ref{fig:DirectDetection} for flavour universal couplings $b_N=\tilde{b}_n=3$.

Notice that the lowest values of $M$ are demanded for $\tpi^0$ decay in Scenario A, and only axial couplings are strictly necessary. In that case only spin dependent are induced with cross section of magnitude comparable to Eq.~(\ref{eq:sigmaSI}) which is below present bounds.

On the other hand, Higgs portal interactions will always contribute with
\begin{equation}
  \sigma_{SI}(\tilde{n}\text{-}N) = \frac{(f_N m_N)^2 (\tilde{f}_n \tilde{m}_n)^2}{\pi m_h^4 f^4}  \mu^2 \simeq 10^{-48}~cm^{-2}\,\, ,
\end{equation}
where $\tilde{f}_n\simeq f_N\simeq 0.3$ is the nucleon matrix element. The predicted cross section is below the neutrino floor.

The presence of effective operators will give additional collider signatures on top of the usual TH signatures~\cite{Burdman:2014zta}. Twin quarks are pair produced and any twin particle will typically escape the detector. Mono-jet~\cite{Aad:2015zva,Khachatryan:2014rra} and mono-photon searches~\cite{Khachatryan:2014rwa} can be safely evaded for $M \gtrsim 1$ TeV. For a more tuned scenario in which the $\tpi$ are heavy enough to decay inside the detector it could be possible to have signatures similar to emerging jets~\cite{Schwaller:2015gea}.

\section{Conclusions}\label{concl}

We have shown that Twin Higgs models provide a natural framework for Asymmetric Dark Matter. They automatically contain the required global symmetries and particles, in particular considering the parallel between baryons and twin baryons.

Indeed, twin baryons with mass $\sim 5$ GeV can be naturally obtained by small $Z_2$ breaking effects which induce a higher twin QCD confining scale. It is also natural to expect the same order of magnitude for the baryon asymmetries in the two sectors. We have studied two scenarios differing in spectrum and interactions, and shown that twin neutrons are typical DM candidates.

Regardless of the twin particle content, higher dimensional operators with effective scale $M\sim O(1-10)$ TeV are expected and often necessary for a successful thermal history. Their introduction provides experimental signatures within current or near future reach, both at Dark Matter direct detection experiments and at the LHC.

\small

\section*{Acknowledgements}
We are deeply thankful to Jack Collins, Duccio Pappadopulo, Maxim Perelstein and Andrea Tesi for valuable discussions and for reading the manuscript. We also thank Nathaniel Craig, Andrey Katz, Simon Knapen, Eric Kuflik,  John March-Russell and Pedro Schwaller for useful discussions. The work of MF is supported by the U.S. National Science Foundation through grant PHY-1316222.

\bibliography{refs}
\bibliographystyle{JHEP}

\end{document}